\documentclass[twocolumn]{aastex701}

\usepackage{multirow}

\newcommand{\Msun}{M\ensuremath{_\odot}\,}
\newcommand{\Lsun}{L\ensuremath{_\odot}\,}

\newcommand{\revone}[1]{#1}

\newcommand{\Nice}{\label{Nice}Université Côte d'Azur, Observatoire de la Côte d'Azur, CNRS, Laboratoire Lagrange, 06000 Nice, France}
\newcommand{\HD}{\label{HD}Astronomisches Rechen-Institut, Zentrum f\"ur Astronomie der Universit\"at Heidelberg, M\"onchhofstr.\ 12--14, 69120 Heidelberg, Germany}
\newcommand{\SAI}{\label{SAI}Sternberg Astronomical Institute, Moscow M.V. Lomonosov State University, Universitetskij pr., 13,  Moscow, 119234, Russia}
\newcommand{\CfA}{\label{CfA}Center for Astrophysics --- Harvard and Smithsonian, 60 Garden Street MS09, Cambridge, MA 02138, USA}
\newcommand{\LU}{\label{LU}Faculty of Physics and Earth System Sciences, Leipzig University, Linnestraße 5, 04103 Leipzig, Germany}
\newcommand{\MSU}{\label{MSU}Faculty of Physics, Moscow M.V. Lomonosov State University, Leninskie gory 1,  Moscow, 119991, Russia}

\def\M{M\,}
\def\N{NGC\,}

\def\U{UGC\,}

\def\HI{H{\sc i}\,}
\def\HII{H{\sc ii}\, }

\def\kms{$\textrm{km~s$^{-1}$ }$}
\def\nb{\textsc{nbursts}}

\newcommand{\SII}{[S~{\sc ii}]}
\newcommand{\OIII}{[O~{\sc iii}]}

\newcommand{\NII}{[N~{\sc ii}]}

\newcommand{\Ha}{H$\alpha$\ }
\newcommand{\Hb}{H$\beta$\ }

\newcommand{\NIIHa}{[N~{\sc ii}]/H$\alpha$\ }
\newcommand{\OIIIHb}{[O~{\sc iii}]/H$\beta$\ }
\begin{document}
\title{MUSE Study of Two Giant Low-surface-brightness Galaxies with Compact Satellites}


\author[0000-0002-4342-9312]{Anna S. Saburova}
\affiliation{\SAI}
\email{saburovaann@gmail.com}
\author[0000-0002-1750-2096]{Damir Gasymov}
\affiliation{\HD}
\affiliation{\SAI}
\email{damir.gasymov@stud.uni-heidelberg.de}
\author[0000-0001-7113-8152]{Ivan S. Gerasimov}
\affiliation{\Nice}
\affiliation{\SAI}
\email{ivan.gerasimov@oca.eu}
\author[0000-0002-4755-118X]{Oleg V. Egorov}
\affiliation{\HD}
\email{oleg.egorov@uni-heidelberg.de}
\author[0000-0002-7924-3253]{Igor V. Chilingarian}
\affiliation{\CfA}
\affiliation{\SAI}
\email{igor.chilingarian@cfa.harvard.edu}
\author[0000-0002-9609-7980]{Fedor M. Kolganov}
\affiliation{\LU}
\email{fkolganov@voxastro.org}
\author[0000-0001-9914-4466]{Anatoly V. Zasov}
\affiliation{\SAI}
\affiliation{\MSU}
\email{}
\author[0000-0001-8427-0240]{Evgenii V. Rubtsov}
\affiliation{\SAI}
\email{evgenii.rubtsov@voxastro.org}
\author[0000-0002-8220-0756]{Anton V. Afanasiev}
\affiliation{\SAI}
\email{anton.afanasiev@voxastro.org}
\author[0000-0002-8297-6386]{Mariia V. Demianenko}
\affiliation{Max-Planck-Institut für Astronomie, Königstuhl 17, 69117 Heidelberg, Germany}
\affiliation{Department for Physics and Astronomy, Heidelberg University, Im Neuenheimer Feld 226, 69120 Heidelberg, Germany}
\email{demianenko@mpia.de}
\correspondingauthor{Anna S. Saburova}
\email{saburovaann@gmail.com}

\begin{abstract}
Giant low-surface-brightness disk galaxies (gLSBGs) are rare objects with disk radii up-to 160 kpc and dynamical masses of an order of up to 10$^{12}$ \Msun. Their very existence challenges currently accepted theories of
galaxy formation and evolution, as it is difficult to build such large, dynamically cold disks through mergers without destroying them. 
We present deep MUSE mosaic observations of two nearby gLSBGs with compact elliptical satellites: UGC~1382, which hosts a globally counter-rotating gaseous disk, and AGC~192040, which does not.  We analyze properties of ionized gas and present spatially resolved kinematics and metallicity maps; as well as stellar population analysis for the central regions of the galaxies. The radial gradients of gas-phase metallicities are flat for both galaxies. 
Our observational data indicate that both galaxies experienced mergers several Gyrs ago. However, the scenarios of the formation of giant disks appear to be slightly different for these two systems. For AGC~192040, we propose the gas accretion from the filament \revone{or from the cooling hot halo gas} followed by the intermediate-mass ratio merger with the companion on a prograde orbit. For UGC~1382, multiple gas-rich mergers with companions on retrograde orbits are preferred by the data.
\end{abstract}



\section{Introduction}

\begin{deluxetable*}{lllll}
\tablecaption{Basic properties of the studied galaxies. \newline 
References: {[1]} NED (\url{http://ned.ipac.caltech.edu}), {[2]} \citet{Hagen2016}, {[3]} \citet{saburovaetal2021}, {[4]} \citet{gereb2018}, {[5]} current paper, {[6]} Hyperleda  \citep[\url{http://leda.univ-lyon1.fr/},][]{Makarov2014}. 
\label{tab1}}
\tablehead{
\colhead{Parameter} & \colhead{UGC~1382} & \colhead{Ref.} & \colhead{AGC~192040} & \colhead{Ref.}
}
\startdata
Names & CGCG 386-054; MCG +00-05-048 &  & 2MASX J09473281+1045083 &  \\
RA (J2000.0) & 01h54m41.0334s & [1] & 09h47m32.8077s & [1] \\
Dec. (J2000.0) & $-00^\mathrm{d}08^\mathrm{m}35.957^\mathrm{s}$ &  & $+10^\mathrm{d}45^\mathrm{m}08.681^\mathrm{s}$ &  \\
Distance (Mpc) & $80$ & [2] & $192.1$ & [4] \\
Scale (kpc arcsec$^{-1}$) & 0.39 &  & 0.93 &  \\
Morphological type & S0 & [2] & Sa & [5] \\
Inclination angle & $46^\circ$ & [2] & $27.7^\circ$ & [6] \\
Major axis P.A. & $58^\circ$ & [3] & $45^\circ$ & [4] \\
\HI mass (\Msun) & $1.7 \times 10^{10}$ & [2] & $5.7 \times 10^{10}$ & [4] \\
Stellar mass (\Msun) & $8 \times 10^{10}$ & [2] & $3.46 \times 10^{10}$ & [4] \\
Rotation velocity (\kms) & $280$ & [2] & $217.4 \pm 13.6$ & [6] \\
\enddata
\end{deluxetable*}
Giant low surface brightness galaxies (gLSBGs), with disk radii extending up to 160 kpc, provide a unique challenge to current models of galaxy formation and evolution. The prototype of this class, Malin~1 was discovered three decades ago \citep{Bothun1987}. Several dozen of gLSBGs have been found since then, but the consensus is that they are rare \citep{saburova2023}. These galaxies deserve special attention because they host the largest known disks and some of the largest concentrations of baryons ($M_{\mathrm{b}} \sim 10^{11}$\Msun), combined with substantial angular momentum and dynamical masses reaching $10^{12}$~\Msun\ within the disk radius.

The formation of galaxies of comparable mass, such as giant ellipticals or brightest cluster galaxies, is usually attributed to multiple major and minor mergers. However, this scenario does not readily explain gLSBGs: retaining their large angular momentum would require infalling satellites to have preferentially aligned orbital spins, which is not expected in $\Lambda$CDM cosmology. While planar structures of satellites have been observed around systems such as M~31 \citep{ibata2013}, Milky Way \citep{Pawlowski2020}, Cen~A \citep{2018Sci...359..534M} and NGC~4490 \citep{Karachentsev2024}, the total baryonic mass of satellites in these cases falls at least an order of magnitude short of that contained in a typical gLSB disk. \revone{For example,  a total stellar mass of the system of  satellites of Cen~A is $\sim 7.5\times 10^7$ \Msun \citep{Muller2021},  whereas  the stellar mass of the LSB spiral arms alone for UGC~1382 is $1.6\times10^{10}$\Msun. The difference will be even higher if we take into account a gas mass since the dSph satellites of Cen~A are gas poor.  }

Cosmological simulations can reproduce gLSBG-like systems \citep{Zhuetal2018, Zhu2023, Kulier2020}, but still fail to capture the full set of observed properties. In particular, simulated analogues often lack the thin, tightly wound spiral structure seen in real gLSBGs, and the number of gLSBGs with the most extreme disk sizes is lower in simulations than in observations.

The formation mechanisms behind the gLSBGs remain an open question, actively debated in the field \citep{Reshetnikov2010, Lelli2010, Kasparova2014, Galaz2015, Hagen2016, Boissier2016,Saburova2018, saburovaetal2021}. 
The following main scenarios have been proposed to explain the formation of gLSBGs: (a) major merger with the fine-tuned orbital parameters \citep{Saburova2018, Zhuetal2018}; (b) 
multiple minor mergers with gas-rich dwarf galaxies \citep{Penarrubia2006}; (c) the formation of gLSBG due to unusual properties of dark matter halo, namely a large radial scale length and low central density -- the sparse dark halo with shallow potential well \citep{Kasparova2014}; (d) the accretion of metal-poor gas from cosmic filament onto a pre-existing elliptical or spiral high surface brightness (HSB) galaxy 
\citep{Saburova2019}. A combination of these scenarios may also contribute. Observational studies based on photometry, H{\sc i}, and optical long-slit spectroscopy suggest that multiple evolutionary channels are at play  \citep{saburovaetal2021, saburova2024}.

A recent development in this context is the discovery of compact elliptical galaxies (cEs) associated with Malin 1 \citep{Reshetnikov2010,  Saburova_proc2021, BustosEspinoza2025}. Compact ellipticals are rare ($\sim$ 0.5\% of the dwarf galaxy population), small ($r_e < 1$~kpc), and dense systems that typically host old, metal-rich stars \citep{1973ApJ...179..423F, 1985AJ.....90.1681B, Chilingarianetal2009, 2021arXiv210309241F}. They are thought to form through tidal stripping of more massive disk galaxies, losing 90–95\% of their stellar mass in the process \citep{2001ApJ...552L.105B, Chilingarianetal2009, 2014MNRAS.443.1151N, ChilingarianZolotukhin2015}. Further inspection of deep imaging data (DECam, Subaru) has revealed additional candidates, reinforcing the possible connection between gLSBGs and cEs.

In this work, we focus on two systems where a gLSBG is found in association with a compact elliptical: UGC~1382 and AGC~192040. We present the first results from our deep mosaic observations of these two gLSBGs obtained with Multi-Unit Spectroscopic Explorer on the VLT \citep[MUSE-WFM,][]{bacon2010}. The main goal is to constrain the formation scenarios of these systems by examining the spatially resolved metallicity distribution of the ionised gas.

The paper is structured as follows. Section \ref{sec:sample} introduces the studied systems. Section \ref{sec:obs} outlines the MUSE observations and data reduction procedures. Spectral analysis and surface photometry are presented in Sections \ref{sec:analysis_data_cubes} and \ref{sec:surfphot}, respectively. The main results are reported in Section \ref{sec:res}, and we conclude with a discussion and summary in Sections \ref{sec:dis} and \ref{sec:sum}.

\begin{figure*}
  \centering

  \includegraphics[width = 0.47\textwidth]{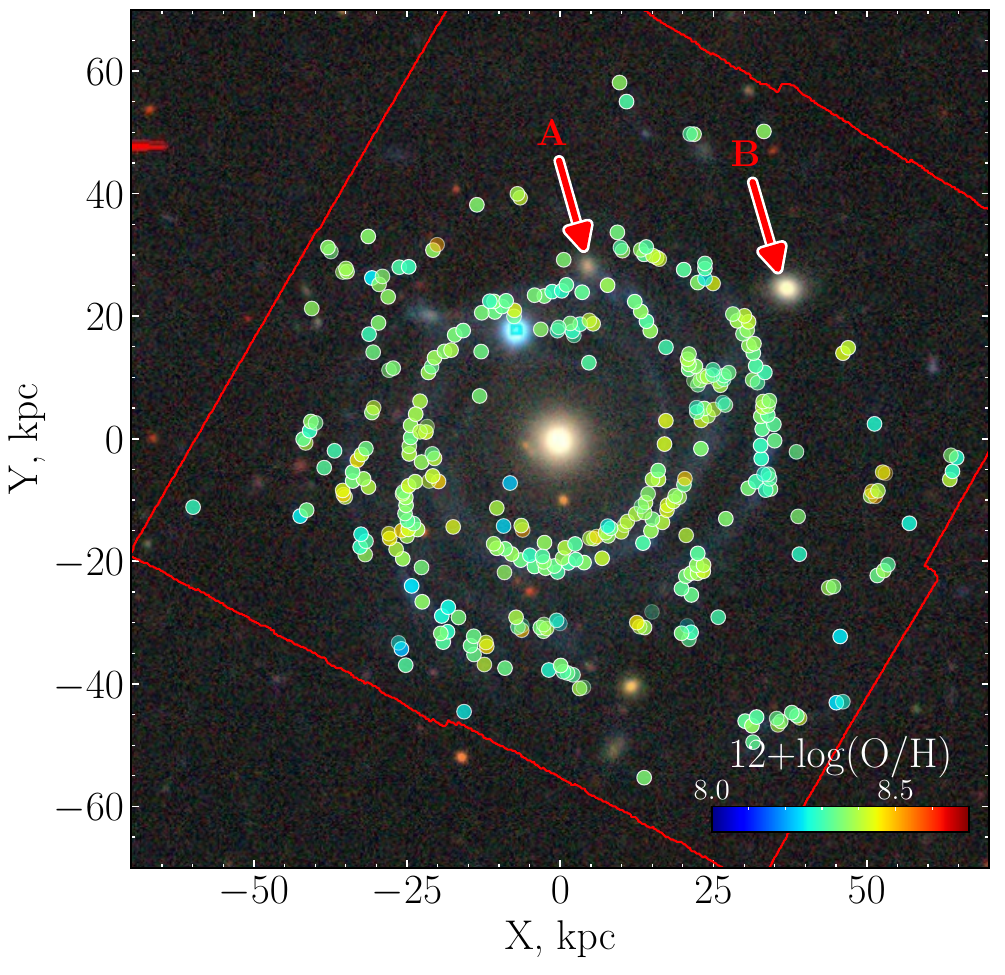}
  \includegraphics[width = 0.47\textwidth]{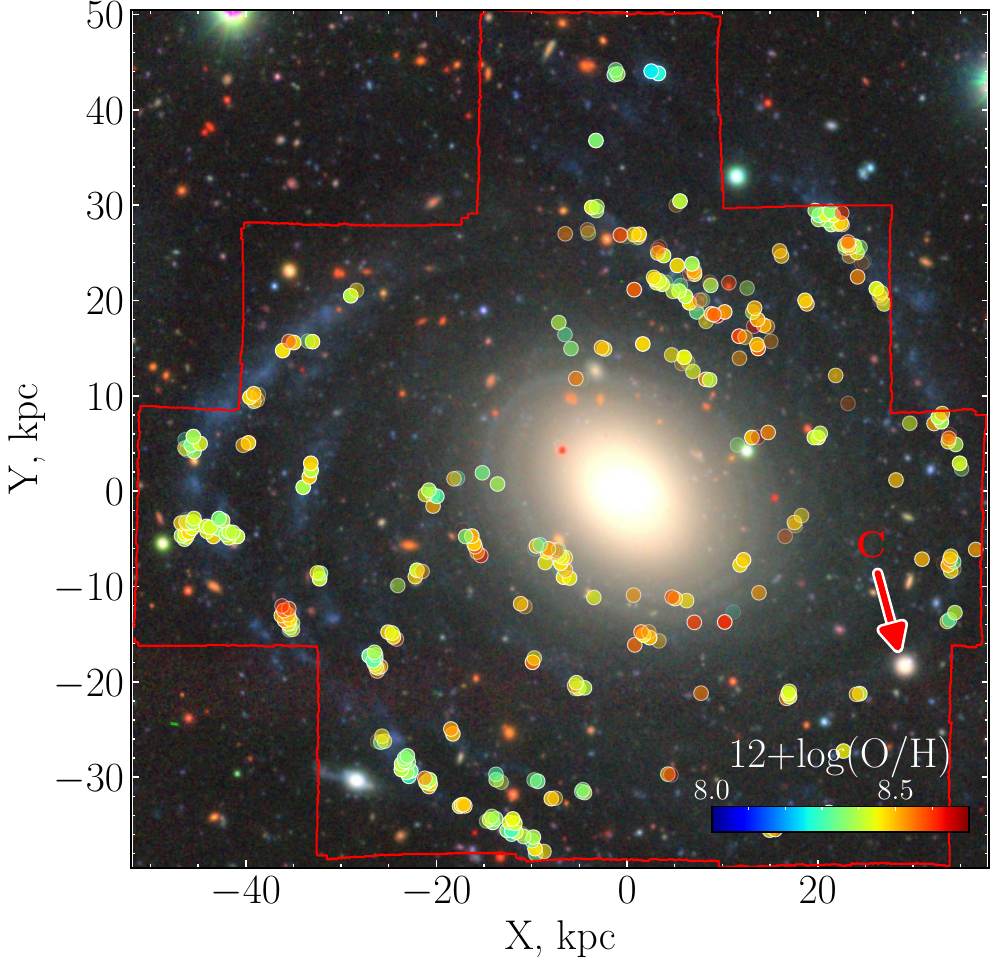}
  \caption{Gas-phase metallicity maps for AGC~192040 (left) ans UGC~1382 (right). The red outline indicates the field of view of the MUSE mosaics. Both maps are based on O3N2 calibration \citep{Marino2013A&A...559A.114M}. Each dot corresponds to a spatial bin; its color represents the oxygen abundance. The transparency of the dots reflects the uncertainty in the metallicity measurement, with more transparent points corresponding to higher uncertainties. The red arrows \revone{with letters} demonstrate the positions of the satellites. For AGC~192040, the cE satellite is \revone{marked as }\revone{(B)}.
  }\label{zmap}
\end{figure*}
\section{Sample}\label{sec:sample}
We investigate two gLSBGs with visually estimated stellar disk radii $R_d\sim80$~kpc hosting cE satellites. Initially considered early-type galaxies, they were re-classified as gLSBGs after their extended disks were detected \footnote{ \revone{Both of them were classified as ellipticals in Hyperleda, see the reclassification and the reference in Table~\ref{tab1}}}. Their main properties are summarized in Table~\ref{tab1}.

These two galaxies share several morphological features: blue low-surface-brightness spiral arms, prominent red bulges, and a characteristic gap between the bulge and spiral structure. Most importantly, both galaxies host compact satellites \citep{Saburova_proc2021} \revone{(B and C in Fig. \ref{zmap})}. \revone{Given that the velocity differences between the considered gLSBGs and the compact ellipticals within the field of view are comparable to the rotation velocities of the gLSBGs, and considering the overall rarity of both galaxy types \citep[see][]{ChilingarianZolotukhin2015, saburova2023}, the chance of their accidental superposition is low. This conclusion is based  not only on the three published cases (UGC 1382, AGC 192040, and Malin 1) but also on the dozen of other systems we have identified (to be published). Such systems are therefore most likely physically associated and evolutionarily connected.  } 

The two galaxies, however, differ in one significant aspect: UGC~1382's giant gaseous disk counter-rotates with respect to the central stellar body \citep{saburovaetal2021}, while in AGC~192040  the two structures co-rotate. This contrast makes their comparative analysis particularly valuable for disentangling the evolutionary processes that produce such systems. 

\citet{Hagen2016} studied {\bf UGC~1382} in detail and argued that a minor merger (scenario b) was the most plausible formation pathway. However, forming a massive gLSB gas+stellar disk with \citep[$3.3 - 7.4 \cdot 10^{10}$~\Msun;][]{Hagen2016, saburovaetal2021} through just one or two dwarf mergers appears unlikely. 

More evidence for a merger occurring in the formation history of gLSBGs comes from the presence of a cE near UGC~1382, confirmed by an SDSS spectrum identified through SDSS spectroscopy and interpreted as a tidally stripped remnant \citep{Gasymovetal2024}. On the other hand, counter-rotating disks can also result from cold gas accretion. Stellar population studies complicate the picture: the results of spectral energy distribution (SED) fitting \citep{Hagen2016} suggest that the gLSB component in UGC~1382 is 4~Gyr older than the central bulge. Such an age difference rules out the possibility of recent cold gas accretion, since the extended stellar disk must have formed long ago. In contrast, \citet{saburovaetal2021} reported a much older bulge, leaving multiple pathways viable. The new MUSE data presented here allow us to refine these interpretations and test competing scenarios for UGC~1382’s evolution.

{\bf AGC 192040} hosts an extended gaseous LSB disk reaching $r=65$ kpc in H{\sc i} and $r=76$ kpc in deep $R$-band imaging \citep{gereb2018}. The H{\sc i} disk is warped, suggestive of tidal interactions within the small group of galaxies dominated by AGC~192040. While no direct evidence of recent gas accretion is detected in the H{\sc i} data \citep{gereb2018}, past accretion cannot be excluded. Both a merger scenario and disk formation driven by unusual dark matter halo properties remain possible.

\section{Observations and data reduction}\label{sec:obs}

We performed deep MUSE observations of UGC~1382 and AGC~192040 (program 110.24DN, P.I.: Saburova).
The UGC~1382 mosaic was observed in October-December 2022 and August-October 2023. The observations took in total 32.6h (11 pointings, 31 observation blocks, hereafter OB). For AGC~192040, the observations were performed in December 2022, January, April, May, December 2023 and January 2024, which took in total of 11.7h (4 pointings, 12 OBs). We demonstrate the data coverage by red contours in Fig. \ref{zmap} on top of optical images of the considered galaxies.

Observations were carried out in wide field mode with adaptive optics (WFM-AO) in a nominal (non-extended) wavelength range. For each OB, we obtained $4\times542$s on-target and $2\times120$s sky exposures following an Object(O)-Sky(S)-O-O-S-O pattern with small offsets and rotation by 90 degrees between each exposure for 10 fields covering the LSB disk of UGC~1382. We performed 3~OBs per each pointing, except the central region, where we used a single OB with $4\times900$s on-target and $2\times120$s on sky. For AGC~192040, we utilized a slightly different observation scheme with 3~OBs per pointing with: S-O-O-O-S pattern for $3\times680$s on-target and $2\times120$s sky exposures without rotation for all 4 fields of the mosaic. Seeing value varied between OBs from 0.8\arcsec\ to 1.4\arcsec.

Data reduction was performed with the \textsc{pymusepipe}\footnote{\url{https://github.com/emsellem/pymusepipe}} package presented and described in \cite{Emsellem2022}. \textsc{pymusepipe} is a wrapper around the MUSE data processing pipeline software (\textsc{MUSE DRS}, \citealt{Weilbacher2020}) with several additional bespoke procedures developed to improve alignment, flux calibration and mosaicking of different exposures to a single data cube. During data reduction with \textsc{pymusepipe}, MUSE DRS was accessed via EsoRex command line recipes to remove instrumental signatures, including bias subtraction, flat-fielding, wavelength calibration, geometric corrections, sky subtraction, absolute flux calibration. We aligned each individual exposure using the available wide-field reference images obtained in the $r$ broad-band filter, with the transmission completely lying within the MUSE wavelength coverage. Namely, for UGC~1382 we used the {\it r}-band images from the \revone{Hyper Suprime-Cam} \citep[][]{2018PASJ...70S...1M} \revone{Subaru} Strategic Program \citep{hsc2019} \revone{(hereafter HSC-SSP)}, and data from the \revone{DESI} Legacy survey\footnote{\url{https://www.legacysurvey.org/}} \citep{Dey2019} for AGC~192040. Using \textsc{pymusepipe}, we created images for each exposure by convolving the preliminary reduced data cubes with the transmission curves of the corresponding filters. Then, the package derived the necessary offsets to minimize the differences between the images, whose reliability was further visually verified. Manual corrections were applied when necessary. Together with the offsets, \textsc{pymusepipe} provides normalization factors by comparing the flux distribution in two images. These factors are then applied to correct the MUSE DRS flux calibration \citep[see][]{Emsellem2022}, for, e.g., small variations due to the slight change in transparency during the OB. In our case, we did not use these additional corrections as their reliability was uncertain for the low surface brightness areas. As a result of data reduction, we obtained mosaicked data cubes covering areas shown by red contour in Fig.~\ref{zmap} and below with a pixel size of 0.2\arcsec, wavelength range 4750--9350\AA, spectral resolution (FWHM of LSF) $\sim 2.4$\AA\ at 7500\AA\ \citep{Bacon2017} and GLAO-corrected PSF. We did not correct individual exposures for differences in PSF before mosaicking. 

\begin{figure*}
  \centering
  \includegraphics[width = \textwidth]{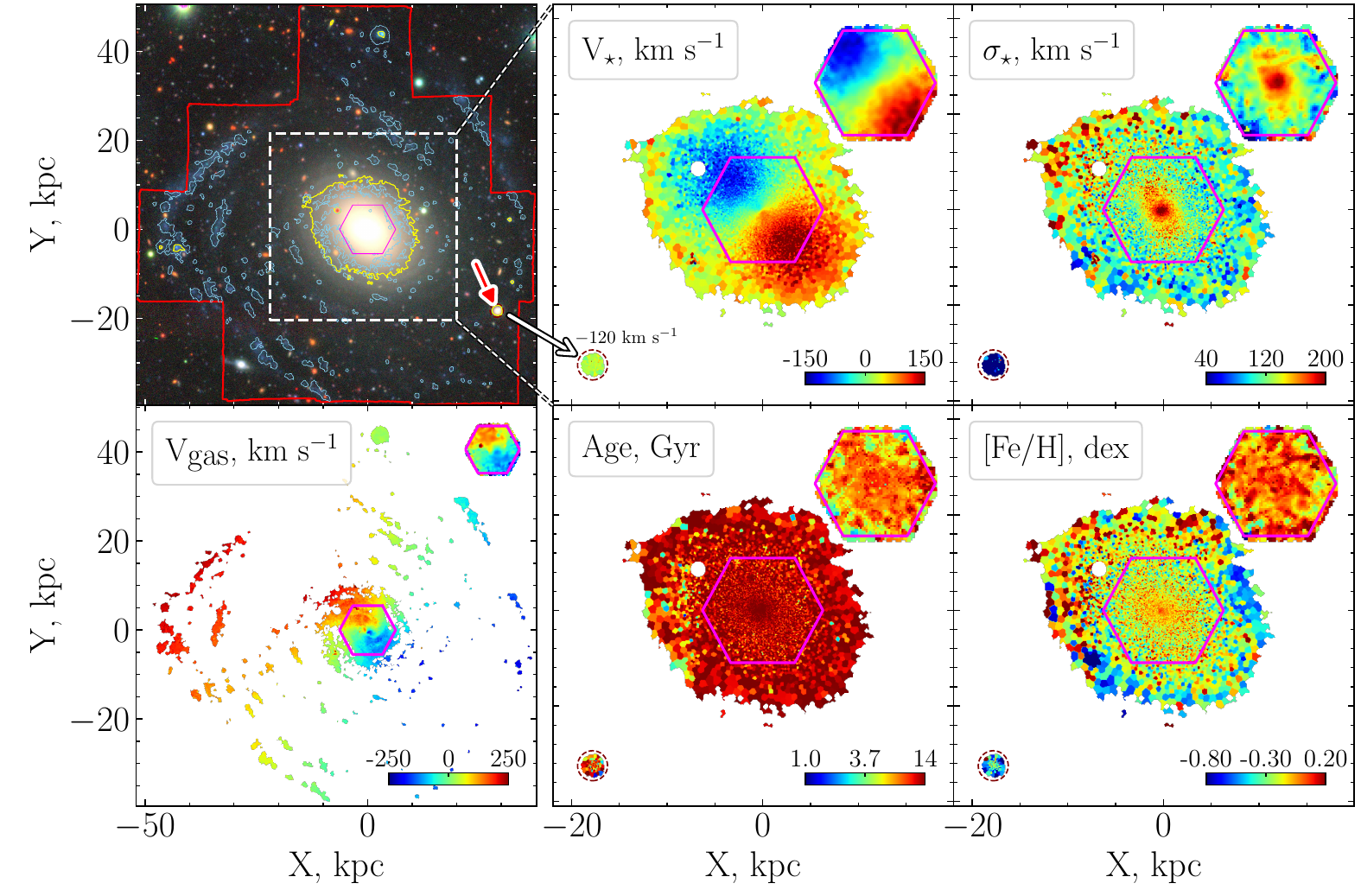}
  \caption{Parameter maps and image of UGC~1382. Top left: HSC-SSP image, with the red outline indicating the field of view of the MUSE mosaics. The magenta hexagon shows the field of view of the MaNGA SDSS spectrum. Light blue and yellow outlines indicate the spatial binning used for fitting emission lines and stellar components, respectively (Sec.~\ref{subsec:nbursts_fit}). Bottom left: Emission line velocity map. In the top right corner, the same map is shown using MaNGA data for the central region (magenta hexagon), with the same color bar and scale (in kpc). Other maps, from left to right and top to bottom: Maps of stellar velocity, stellar velocity dispersion, and SSP-equivalent luminosity-weighted age and metallicity for the central region of the galaxy. In each plot, the top right corner shows the corresponding MaNGA-based map for the central region (same color bar and scale in kpc), and the bottom left corner shows a cutout highlighting the companion (with a velocity shift of $-120$ km~s$^{-1}$).} \label{maps_u1382}
\end{figure*}

\begin{figure*}
  \centering
  \includegraphics[width = \textwidth]{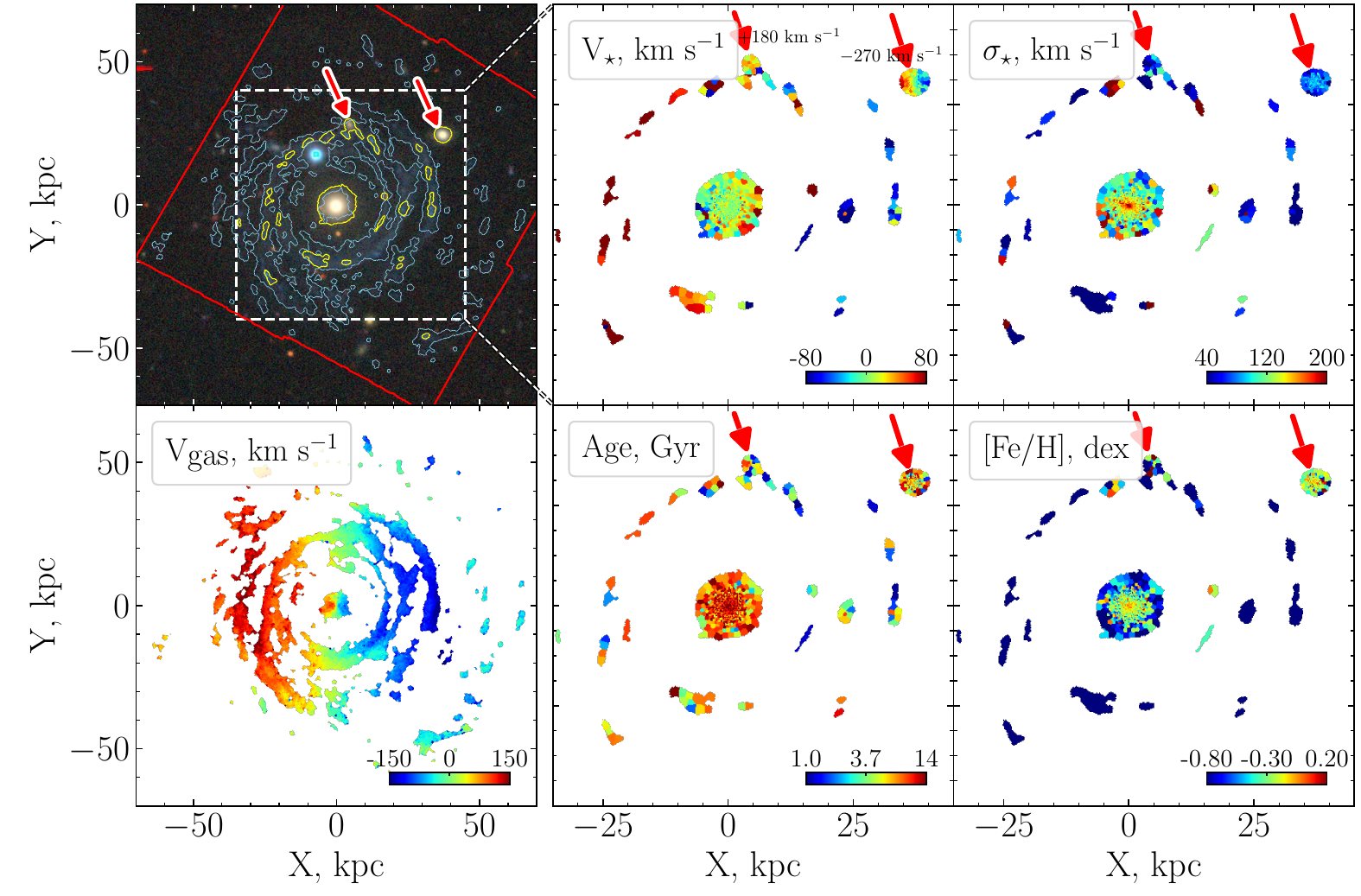}
  \caption{Parameter maps and image of AGC~192040. Top left: \revone{DESI} Legacy Survey image, with the red outline indicating the field of view of the MUSE mosaics. Light blue and yellow outlines indicate the spatial binning used for fitting emission lines and stellar components, respectively (Sec.~\ref{subsec:nbursts_fit}). Bottom left: Emission line velocity map. Other maps, from left to right and top to bottom: Maps of stellar velocity, stellar velocity dispersion, and SSP-equivalent luminosity-weighted age and metallicity for the galaxy. The two stellar companions --- located above the center and in the top right corner --- have systemic velocities relative to the galaxy of $+180$~km~s$^{-1}$ and $-270$~km~s$^{-1}$, respectively (the stellar velocity bins corresponding to these companions were shifted accordingly).} \label{maps_agc}
\end{figure*}

\section{Analysis of the mosaic data cubes}
\label{sec:analysis_data_cubes}

We analyze the data cubes of UGC~1382 and AGC~192040 (MUSE mosaics; Section~\ref{sec:obs}) following the approach of \citet{saburova2024}, adapted from long-slit to integral-field data. Briefly, we design task-specific spatial binning schemes (Section~\ref{subsec:binning}) and fit each bin independently with \nb{} to obtain stellar and gaseous properties (Section~\ref{subsec:nbursts_fit}). Gas-phase diagnostics are then measured from fitted emission line fluxes and examined as a function of galactocentric distance (Section~\ref{subsec:gas_metallicity}).

\subsection{Spatial binning schemes}
\label{subsec:binning}

Different goals require different treatments of the cubes, so we adopt three binning schemes targeting: (i) stellar population properties in the central regions and satellite galaxies; (ii) gas kinematics across the full field of view; (iii) gas-phase metallicity in clumps of the LSB disk. We retain this numbering below.

For each galaxy, we compute signal-to-noise (S/N) ratio maps from the original data cubes using: (i) the median continuum flux and its error in the rest-frame 5050--5150~\AA\ window; (ii) the continuum-subtracted total flux and error in a narrow band centered on redshifted \Ha line; and (iii) an analogous map for \Hb line. Foreground/background sources are masked via a cross-match with \textit{Gaia} DR3 \citep{GaiaCollaboration2023A&A...674A...1G} and by visual inspection of spectra.

We reject low-S/N spaxels with a conservative threshold ($\mathrm{SNR}_{\min}\!\sim\!1$--$1.5$). The remaining regions must contain at least 100 and 25 spaxels for schemes (i) and (ii), respectively, while for (iii) were used deblending procedure provided as a part of \textsc{photutils} package to distinguish individual \HII regions. Within each accepted region for (i) and (ii) schemes, we apply Voronoi binning \citep{Cappellari2003MNRAS.342..345C}, constrained to the region footprint, to reach an approximately uniform target S/N $\approx 10$. This pipeline provides a sufficient number of high-quality bins for both stellar and gas-phase analyses.

\subsection{Full spectrum fitting}
\label{subsec:nbursts_fit}

We pass the mosaics and the binning schemes to \nb{} \citep{Chilingarian2007a,Chilingarian2007b,ChilingarianKatkov2012}, which performs full-spectrum fitting with E-MILES simple stellar population (SSP) models \citep[][FWHM~$=2.5$~\AA]{Vazdekis16} for the stellar component and a set of emission lines within 4800--6800~\AA\ (rest-frame). Telluric regions are masked to avoid residual sky-subtraction noise.

For each bin we derive: SSP-equivalent age and metallicity; stellar and gas kinematics (radial velocities and velocity dispersions); and emission line fluxes (Gaussian profiles with constraints on flux ratios for doublets and for the Balmer series). The parameters are determined sequentially for the three schemes: (i) stellar properties over 4700--8700~\AA\ in rest-frame, including the Ca-triplet region; (ii) gas kinematics over 4800--6800~\AA, then used as initial guesses; (iii) refined emission line fluxes for BPT classification \citep{BPT} and gas-phase metallicity. The \nb{} results are used to construct spatial maps of all model parameters, shown in Figures~\ref{maps_u1382} and~\ref{maps_agc} for UGC~1382 and AGC~192040\footnote{\revone{The FITS-files with the maps are available via \url{https://zenodo.org/records/17829726}}}, and discussed in Section~\ref{sec:res}.

Our emission line data are characterized by the following limiting H$\alpha$ flux: \revone{$\sim~2\times$10$^{-18}$~$\rm erg\ s^{-1}\ cm^{-2}\ arcsec^{-2}$ for both galaxies.} 

\subsection{Gas-phase metallicity measurements}
\label{subsec:gas_metallicity}

\begin{figure*}
  \centering

  \includegraphics[width = 0.48\textwidth]{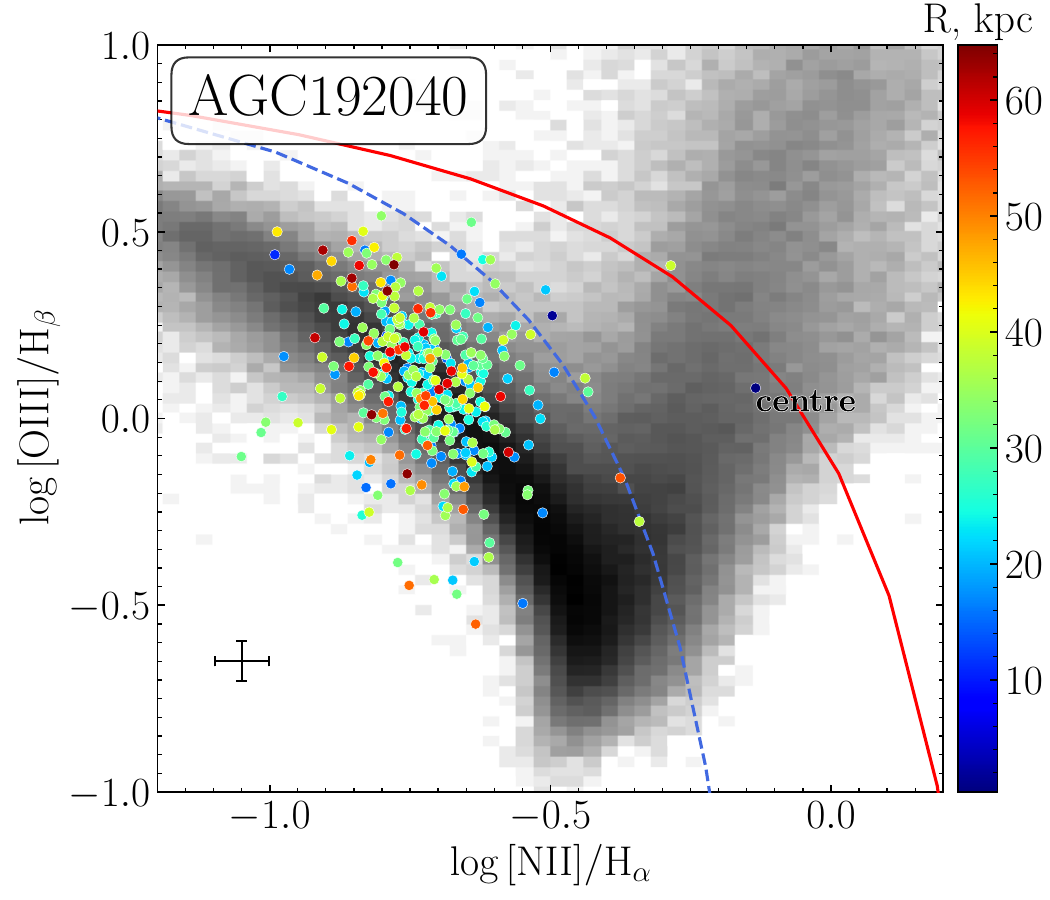}
  \includegraphics[width = 0.48\textwidth]{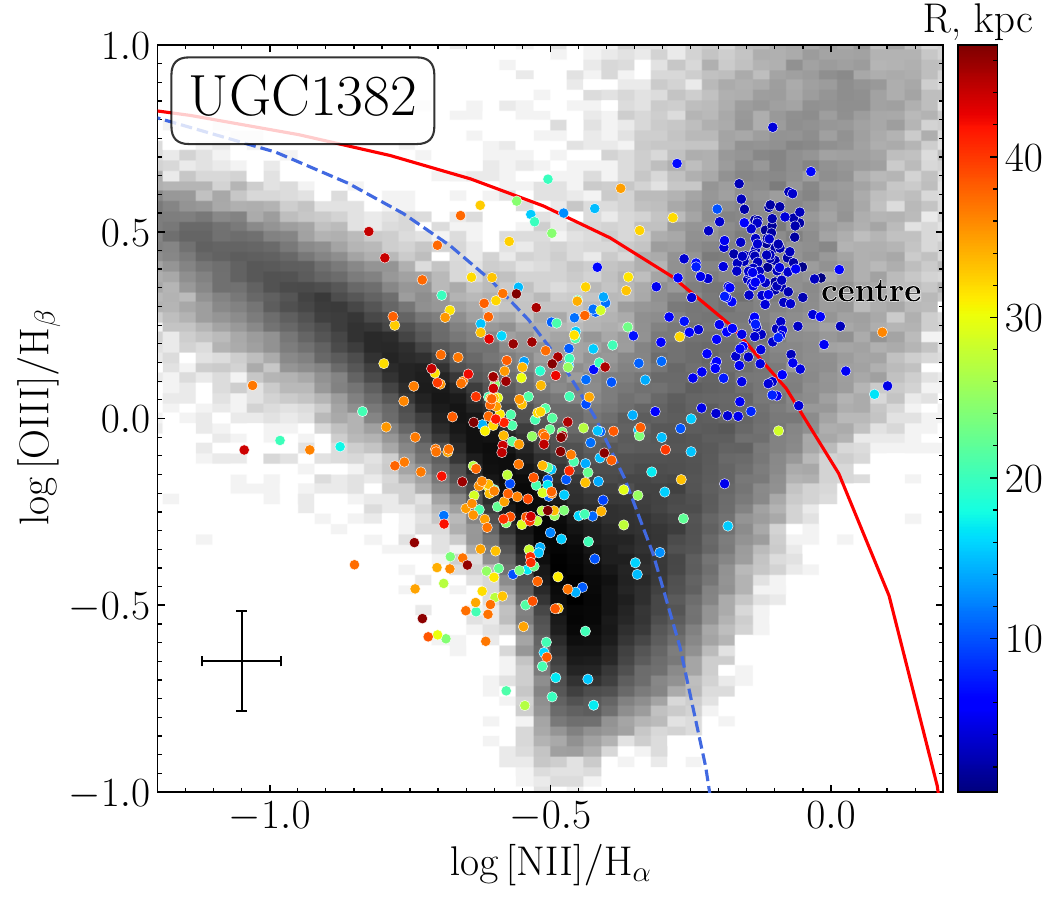}
  \caption{BPT diagrams for AGC~192040 and UGC~1382. Symbols are color-coded corresponding to their galactocentric distance, the cross in the bottom-left corner represent the median uncertainties of the parameters. The blue dashed line represents the demarcation line introduced by \citet{Kauffmann2003}, which distinguishes between the star-forming and the so-called composite regions. The additional demarcation lines are taken from \citet{Kewley2001}. Density plots of emission line measurements from the RCSED database
    \citep[\url{http://rcsed.sai.msu.ru},][]{Chilingarian2017ApJS..228...14C} are shown in grey.}\label{fig:bpt}
\end{figure*}

\begin{figure}
  \centering
  \includegraphics[width = \linewidth, trim= {0 .5cm 0 0}, clip]{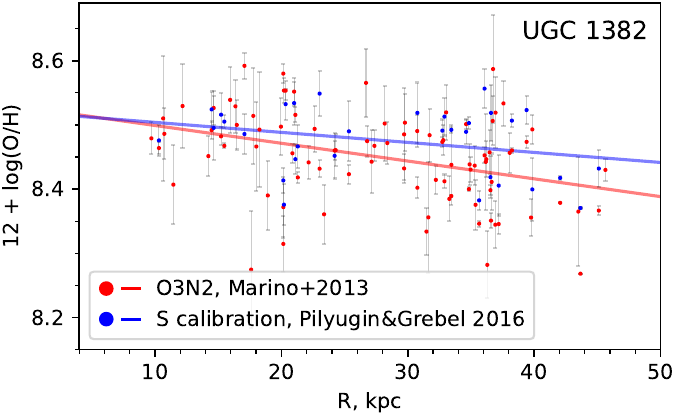}
  \includegraphics[width = \linewidth]{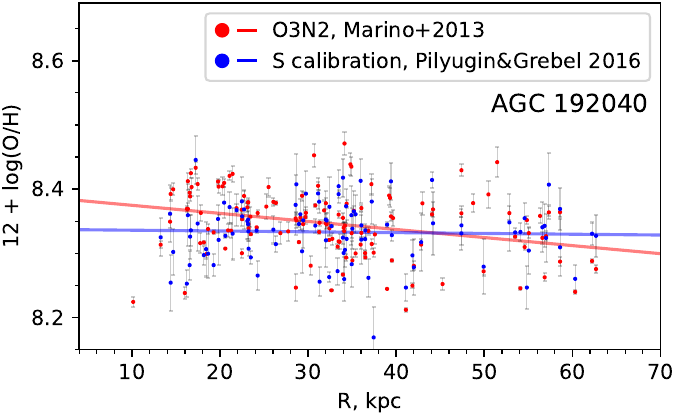}
  \caption{Radial metallicity distribution for UGC~1382 and AGC~192040. Solid lines are a linear regression of the data points. \revone{The  solar metallicity corresponds to the top of the frame being equal to 12 + log(O/H)=8.69 \citep{Asplund2009ARA&A..47..481A}.}}\label{fig:zprofile}
\end{figure}

We measured the oxygen abundance (a proxy of gas phase metallicity) to search for signatures of recent gas infall or ongoing gas accretion from cosmic filaments. To this end, we selected individual \HII regions based on the corresponding binning scheme (iii). Using the emission line fluxes obtained from the \nb{} fitting procedure and corrected for extinction, we checked whether each region fell within the photoionization zone on the diagnostic BPT diagrams (see Fig. \ref{fig:bpt}). Regions lying above the demarcation curve from \citet{Kewley2001} were excluded from the analysis. For the remaining \HII regions, we estimated metallicities using two methods: the O3N2 calibration \citep{Marino2013} and the S-calibration \citep{Pilyugin2016}. The first approach utilises measurements of \OIIIHb{} and \NIIHa{} line ratios, while the latter uses the \OIII{}, \SII{}, and \NII{} lines relative to \Hb{}. We constructed radial metallicity profiles for both galaxies, as shown in Fig.~\ref{fig:zprofile}. We give the values of metallicity and its radial gradient obtained with the two different metallicity calibrations in Table ~\ref{tab:metallicities}.

\begin{deluxetable*}{ccc}
   \tablecaption{Metallicity parameters measured with O3N2 and S-cal methods}
   \label{tab:metallicities}
   \tablehead{
      \colhead{Parameter} & \colhead{UGC~1382} & \colhead{AGC~192040}
   }
   \startdata
   \multicolumn{3}{c}{\textbf{O3N2 calibration}} \\
   12 + log(O/H) & $8.53 \pm 0.02$ & $8.38 \pm 0.01$ \\
   O/H grad., dex kpc$^{-1}$ $\times 10^{-3}$ & $-2.8 \pm 0.5$ & $-1.13 \pm 0.06$ \\
   \hline
   \multicolumn{3}{c}{\textbf{S calibration}} \\
   12 + log(O/H) & $8.52 \pm 0.02$ & $8.33 \pm 0.01$ \\
   O/H grad., dex kpc$^{-1}$ $\times 10^{-3}$ & $-1.6 \pm 0.5$ & $-0.1 \pm 0.2$ \\
   \enddata
\end{deluxetable*}

\section{Surface photometry}\label{sec:surfphot}

We also required the structural parameters of the galaxies’ main stellar components. For UGC~1382 we adopted {\it g}-band values from Kolganov et al. (in prep.) \revone{based on HSC-SSP data}, where the fit contained a S\'ersic bulge component and two exponential disks with the following central surface brightnesses and radial scalelengths: $\rm (\mu_0)_{HSB}=21.19\pm0.10$ mag~arcsec$^{-2}$, $\rm h_{HSB}=3.73\pm0.13$ kpc; $\rm (\mu_0)_{LSB}=25.51\pm0.15$ mag~arcsec$^{-2}$, $\rm h_{LSB}=26.39\pm2.53$ kpc. For AGC~192040, we performed surface photometry in the current paper for the {\it r}-band image of this galaxy from the \revone{DESI} Legacy Survey following the approach from Kolganov et al. (in prep.), which we briefly describe below.

First, we adopted the extraction of the 1D surface brightness profile of the galaxy using the {\sc ellipse} task in the {\sc python photutils} library \citep{photutils} as described in \citep{saburova2023}. After that, we modeled the profile. The model consists of the following components: the S\'ersic's profile, the exponential disk and two Gaussian rings. We implemented the rings into the model to account for irregularities in the 1D profile caused by the isophotal fit along the spiral arms. They, therefore, are not counted as structural components of the galaxy and are not changed throughout the last two steps of the fitting procedure. It was carried out as follows: (i) initial least-squares fit with the model, consisting of S\'ersic' profile, exponential disk, and two Gaussian rings; (ii) S\'ersic' and exponential profiles parameter estimation using Maximum-Likelihood (ML) method; (iii) error estimation with MCMC simulations around the ML-estimate.

\begin{figure}
  \centering
  \includegraphics[width = \linewidth]{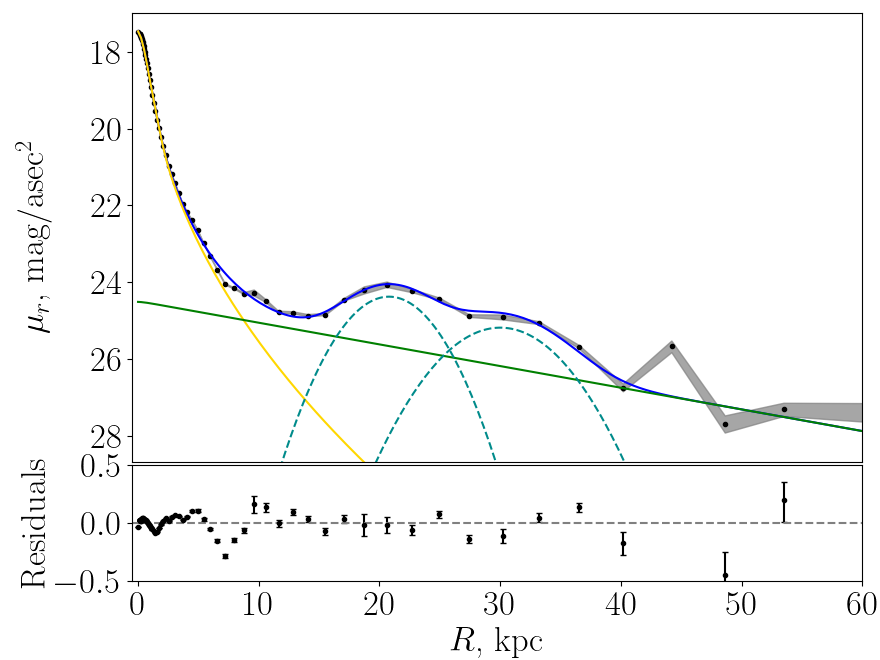}
  \caption{ The {\it r}-band 1D photometric profile decomposition of AGC~192040 (top panel).The black circles with shaded error bars give the observed profile. The blue line gives the total model. Green and yellow lines demonstrate the contributions of the disk and S\'ersic bulge, respectively. Dashed lines show the two Gaussian rings. The bottom panel shows the residuals of the model.}
  \label{fig:agc_1d_decomp}
\end{figure}

In Fig. \ref{fig:agc_1d_decomp} we show the results of the surface brightness profile decomposition. The resulting exponential disk radial scalelength $h=18.48\pm 1.14$ kpc and the central surface brightness $(\mu_0)_r=24.47\pm0.10$ mag./arcsec$^2$ satisfy the gLSBGs selection criterion adopted in \citep{saburova2023}.
\

\section{Results of spectral data analysis}\label{sec:res}

In Fig. \ref{maps_u1382} (bottom left panel), we show the velocity map of ionized gas for UGC~1382, where we combine the results of our MUSE observations with those from MaNGA SDSS \citep{manga} (central region). For the best comparison, we fitted MaNGA data in the same spectral fitting setup as we did for our MUSE observations. The results based on MaNGA agree well with those for MUSE, which additionally validates the analysis performed in the current paper.

As one can see from Fig. \ref{maps_u1382} (bottom left panel), the positional angle of the major axis of the ionized gas velocity map changes with radius. The kinematically decoupled inner region could be the bar. The indication for the presence of the bar also follows from the \textsc{galfit} \citep{galfit} modeling of the {\it r}-band image of this galaxy, the residual of the model that includes bulge and disk shows a prominent small bar (see Fig. \ref{galfit}). The inset shows the HST F814W residual image for the innermost part of the galaxy, where we see a strong dust lane well aligned with the bar.

Kinematic major axes of stars and gas are offset by $\sim$ 180~deg in position angle (see Fig. \ref{maps_u1382}, top middle panel). Thus, we confirm the presence of gaseous counter-rotation with respect to stars that was discovered earlier by \cite{saburovaetal2021}. The stellar velocity dispersion (top right panel) has a peak of 200 \kms in the centre. Stellar population in the bulge is old (T $\sim13$~Gyr) and metal-rich ([Fe/H]$\sim0$~dex), so as the inner part of the disk (see Fig.\ref{maps_u1382}, bottom middle panel). We found the younger ages (6~Gyr) in several compact regions of the disk. However, these estimates correspond to a low signal-to-noise ratio in the low brightness regions. 

In Fig.~\ref{zmap}, we demonstrate the gas metallicity maps for AGC~192040 (left-hand panel) and UGC~1382 (right-hand panel) in regions ionized by massive stars. The metallicity shows almost flat radial gradient for both galaxies, which may evidence the mixing of gas. The same conclusion follows from the radial profile of metallicity demonstrated in Fig.~\ref{fig:zprofile}  (also see the values of radial gradient in Table~\ref{tab:metallicities}). Its mean value is only slightly sub-solar ($12 + \log(\mathrm{O}/\mathrm{H}) = 8.5$) for UGC~1382. At the same time a prominent feature that catches eye is the area with lower metallicity ($12 + \log(\mathrm{O}/\mathrm{H}) = 8.3$) on top of image of UGC~1382 \revone{(+3,+44 kpc)} which coincides with the blue star-forming clump (see the bright blue point on top of right panel of Fig.  \ref{zmap}.) Note that a local region with the metallicity \revone{$12 + \log(\mathrm{O}/\mathrm{H}) = 8.25$ (based on the T$_e$ method)} was earlier found near the center of the dwarf LSB galaxy Ark 18 with the morphology resembling that of gLSBGs \citep{Egorova2021}.
Likewise, \revone{a bright region has been observed near the center of Malin 1 \citep{Johnston2024} which may also represent an infalling star-forming gas clump. Although the clumps in Malin 1 and Ark 18 exhibit significantly higher surface brightness (the mean {\it r}-band surface brightness of the  feature under discussion in UGC~1382 is $23.15$~mag./arcsec$^2$), such anomalies may share a common origin and provide evidence for the infall of gas clumps that locally triggers star formation within galactic disks.}

Figure \ref{maps_agc} presents similar maps for AGC~192040. The velocity map of ionized gas shows that the position angle of the major kinematic axis does not change significantly with the radius in the inner part of the galaxy, while the outer disk is warped, in agreement with \HI observations \citep{gereb2018}. No counter-rotation between gas and stars is detected. Gas metallicity in AGC~192040's disk \revone{is 0.15 dex lower than} that of UGC~1382, being equal to $12 + \log(\mathrm{O}/\mathrm{H}) = 8.3\,\text{--}\,8.4$.

The stellar velocity map of AGC~192040 shows that the reddish clump in the north (\revone{A} in Fig.~\ref{zmap}) has a stellar velocity about 100~\kms higher than the gaseous disk. This clump also exhibits high velocity dispersion and a stellar age of $\sim$5 Gyr, suggesting it is a diffuse satellite or a remnant of one, distinct from the compact satellite \revone{(B)}.

\revone{We carefully inspected the data cubes for both galaxies, examining the spectra and all available information for each detected source. This analysis confirmed that all objects other than those indicated by the arrows or explicitly discussed in the text are either foreground stars or background galaxies. This includes the curious blue source with a tail located near the bottom of the disk of UGC 1382, which we identify as a background galaxy with a redshift of $z=0.0797$.  }

Figure \ref{fig:bpt} presents the BPT diagrams for both galaxies, with points color-coded by galactocentric radius.
In UGC~1382, the inner points lie above the demarcation line defined by \cite{Kewley2001}, suggesting a possible \revone{low-mass} AGN contribution \revone{and a hint at low-ionization nuclear emission-line region (LINER)}. The inner region of AGC~192040 falls within the composite zone, consistent with a possible, though not conclusive, \revone{low-luminous} AGN \revone{ (LINER)}. 
However, \texttt{unTimely}~\citep{Meisner_2023} catalog of co-added WISE and unWISE photometry does not reveal the properties of AGN (neither correlation in the $W1$ (3.4~$\mu m$) and $W2$ (4.6~$\mu m$) photometric bands over 10 years, nor AGN color according to $W12-W3$ diagram~\citep{2010AJ....140.1868W}), both in UGC~1382 and AGC~192040. The ``blue'' $W12$ color ($W1-W2<0$) of UGC~1382 points to a galaxy-dominated center (at least in the WISE PSF of~6.5'')\revone{, which is generally consistent with LINER~\citep{2017FrASS...4...34M}}.

Most outer regions of AGC~192040 lie within the star-forming (photoionization) domain, indicating no clear evidence for shock-excited gas; the same holds for the outermost radii of UGC~1382.

\section{Discussion}\label{sec:dis}
\subsection{UGC~1382}

 \begin{figure}
  \centering
   \includegraphics[width = \linewidth]{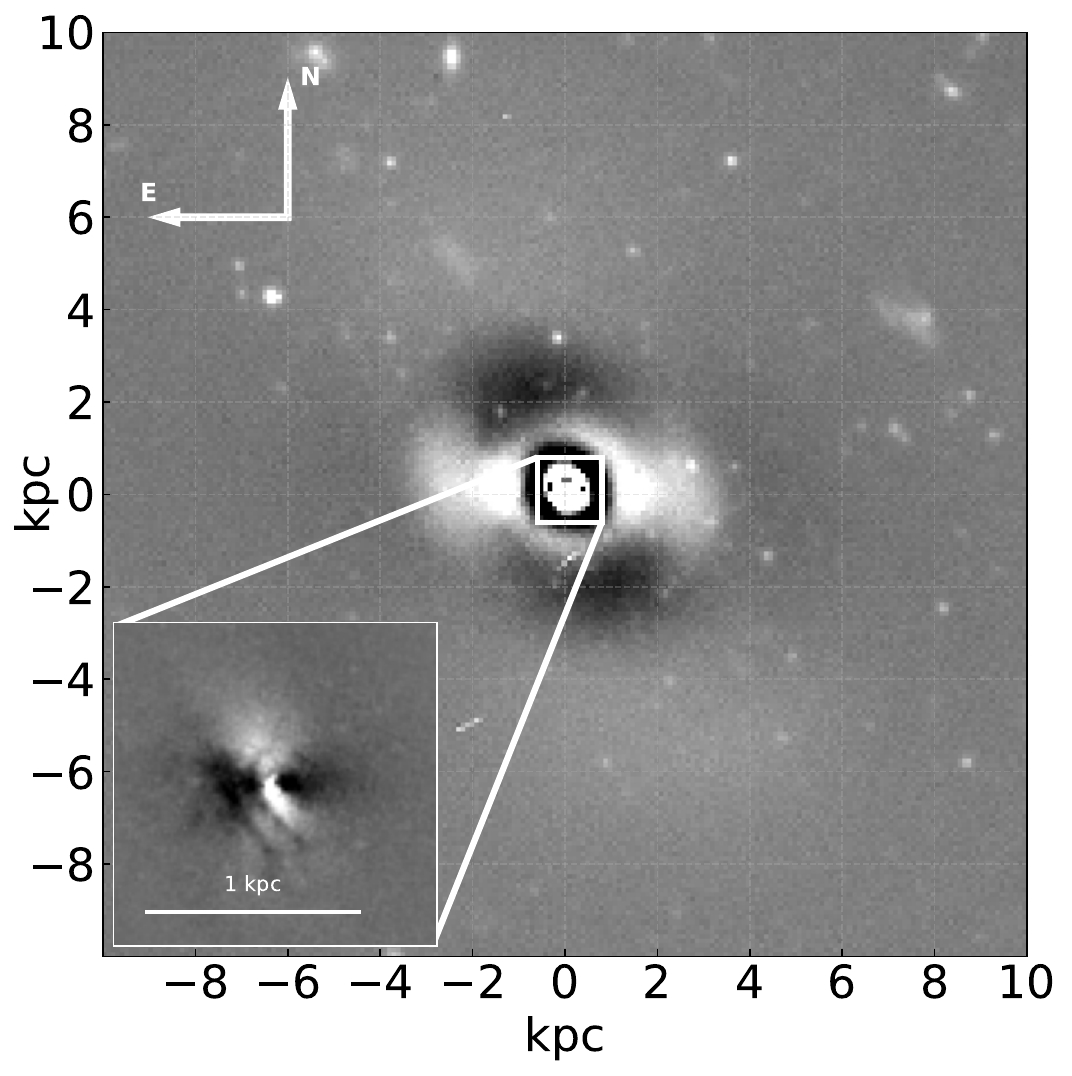}
  \caption{Residual of 2d-photometry model from {\it r}-band HSC-SSP (main) and HST F814W (inset) images of UGC~1382. The length of the bar is about 8 arcsec.}\label{galfit}
 \end{figure}

\begin{figure}
  \centering
  \includegraphics[width=\linewidth]{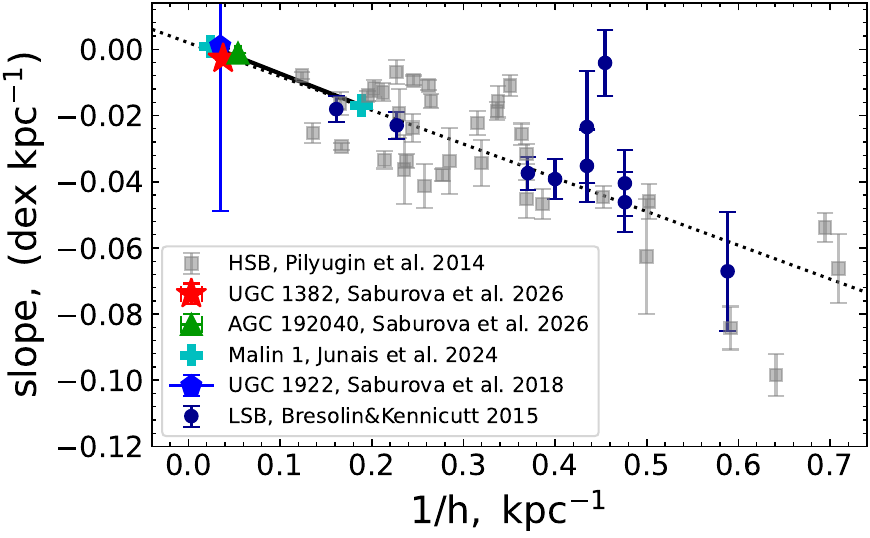}
  \caption{The relation between the metallicity radial gradient and the disk scale length. UGC~1382 and AGC~192040 are shown by the star and triangle. The grey squares give the position of the sample of HSB galaxies (slopes and scale lengths are taken from \citealt{Pilyugin2014a} and  \citealt{Pilyugin2014b} respectively). Blue circles demonstrate the LSB sample from \citet{BresolinKennicutt2015}. We also show for a comparison the position of two gLSBGs: UGC~1922 \citep{Saburova2018} and Malin~1 \citep{Junais2024} as blue pentagon and cyan plus, respectively. Note that Malin~1 has two disks. The black dashed line is eq.~1 adopted from \citet{BresolinKennicutt2015}.}
  \label{fig:gradh}
\end{figure}

\begin{figure}
  \centering
  \includegraphics[width=\linewidth]{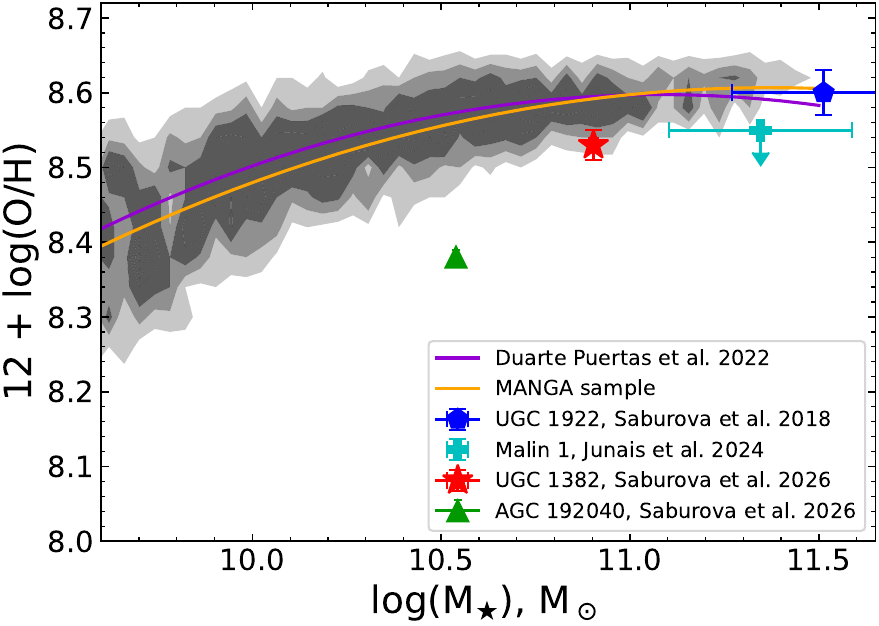}
  \caption{Mass-metallicity relation. UGC~1382 and AGC~192040 are shown by a star and a triangle, respectively.  The stellar masses are taken from \citep{Hagen2016, gereb2018}. For comparison, two other gLSBGs are shown: Malin~1 (\citealt{Junais2024}) (cyan plus) and UGC~1922 \citealt{Saburova2018} (blue pentagon).   
   We overlay the relation obtained for the local Universe (z = 0) from \citet{DuartePuertas2022} (violet), which uses empirical calibrations, that agree well with the MANGA sample metallicities determined with the S2 calibration \citet{Pilyugin16} (superimposed background 2D source density distribution and its 2nd-degree polynomial mean approximation shown by the orange line).  }
  \label{fig:ohmstar}
\end{figure}

To interpret the chemical abundance distribution in UGC~1382, we compare the measured radial gradients with those observed in normal disk galaxies in metallicity gradient -- disk size (Fig.~\ref{fig:gradh}) and stellar mass -- metallicity (Fig.~\ref{fig:ohmstar}) relations. For the latter, we adopt metallicities interpolated to the galaxy centers, since the gradients are shallow and the innermost regions fall within the AGN zone, preventing a direct metallicity measurement there. The stellar mass for UGC~1382 is taken from \citep{Hagen2016}.  In these figures, we compare our data with HSB and LSB galaxies from the literature.

As one can see from Fig. \ref{fig:ohmstar} average gas metallicity of UGC~1382 is slightly lower than expected for its stellar mass for `normal' discy galaxies. In principle, it can indicate that the gas of UGC~1382 was diluted by the fresh supply of metal-poor gas, e.g., accreted from the gas-rich satellites. The observed cE can be a leftover of the satellite galaxy that brought material for the formation of the disk of UGC~1382. So is the metal-poor blue blob seen in the periphery of the disk (see Fig.\ref{zmap}). The galaxy’s central region shows a complex structure: a thick dust lane aligned with the bar, and another stellar component oriented perpendicularly, which may trace a past merger (see the inset of Fig. \ref{galfit}). Moreover, the global gaseous counter-rotation relative to the stars strongly supports an external origin of the disk material. At the same time, the velocity field of UGC~1382 does not show any regions in the disk with differing velocities that would signal recent accretion, implying that the system has had time to settle after these events.

The metallicity gradient of UGC~1382 is in agreement with the radial scalelength of its giant disk, giving evidence that the evolution of the disk size and metal enrichment might be similar to that of the other discy galaxies.

\subsection{AGC~192040}

Unlike UGC~1382, AGC~192040 is an \HI-excess galaxy, with an \HI mass nearly twice its stellar mass. A deep {\it r}-band image shows no stellar shells or streams expected from a merger \citep{gereb2018}. The warp of the \HI disk may result from tidal interaction within its small group. Despite its large gas reservoir, AGC~192040 has an unusually low molecular-to-neutral gas fraction, pointing to inefficient molecular gas formation and, consequently, inefficient star formation \citep{lee2014}, similar to what is observed in the prototypical gLSBG Malin~1 \citep{galaz2024}.

 AGC~192040 also exhibits a significantly lower gas-phase metallicity than expected for its stellar mass taken from \citet{gereb2018} (see Fig.~\ref{fig:ohmstar}). This discrepancy may indicate recent accretion of gas from metal-poor, gas-rich satellites or a cosmic filament (which agrees with the galaxy's observed \HI excess). However, there are no clear traces of recent gas accretion in \HI kinematics of this galaxy \citep{gereb2018}. 
 
 Another possible reason for the reduced metallicity of the observed gas in this galaxy (not excluding the first one) is the less advanced stage of the chemical evolution of its disk, when a significant portion of the initially un-enriched gas had not yet had time for transition into stars due to the low efficiency of star formation (see also Sec.~\ref{subsec:oxygen_yield}). This applies to galaxies in which the mass of gas constitutes a significant fraction of the baryonic mass (at least in the outer disk regions). In this respect, the disks of AGC~192040 and UGC~1382  may be similar to the disks of irregular galaxies characterized by high gas content: they also lie below a big sample of starforming galaxies on the $\log (\text{O}/\text{H}) - \log \text{M}_* $ sequence \revone{although for lower stellar masses} \citep[see Fig.~8 in][]{lee2006}. However, this conclusion applies only to integral estimates of O/H and $M_*$. 
 
 At the same time, the radial gradient of metallicity is not in conflict with that expected for the disk size estimated in the current paper (see Fig. \ref{fig:gradh}). Thus AGC~192040 still follows the universal stellar surface mass density -- metallicity relation \citep{BresolinKennicutt2015}.

Finally, the morphology of AGC~192040 provides further clues. The spiral arms of the galaxy demonstrate  \revone{the presence of  structural features which may be described as the}  long, straight segments. \revone{They are  on the upper-right arm (two long, linear pieces), on the right inner arm (two shorter straight sections), and on the upper-left arm (one straight segment near the foreground star).} Such “rows” are \revone{rather }common in spiral galaxies and are \revone{possibly}  linked to non-stationary processes that create nonlinear compression waves in the gas–star disk \citep[see][and references therein]{2019JPhCS1203a2062B}. In AGC~192040, the most likely driver of these perturbations is the compact satellite currently projected onto the disk. \revone{The observed  connection between the "rows"  in spiral arms and the interaction was noted in \citet{2001ARep...45..841C}.}

\subsection{The effective oxygen yield} 
\label{subsec:oxygen_yield}
Measurements of gas-phase metallicity, combined with estimates of gaseous and stellar surface densities, can provide valuable insights into the chemical evolution of galactic disks. They can help determine whether a galaxy has experienced episodes of gas accretion, either from gas-rich satellites or cosmic filaments, or its evolution proceeds under conditions akin to the so-called closed-box model, in which no significant inflow or outflow of gas occurs. 

A particularly useful diagnostic in this context is the calculation of the oxygen yield. The oxygen mass fraction $\rm Z\approx 12 (O/H)$ is related to the gas-to-total baryonic density ratio ($\mu$) as well as to the total mass fraction returned to the interstellar medium through stellar evolution ($r$), including both processed and unprocessed material. According to (\citealt{Searle1972}; \citealt{Edmunds1990}; \citealt{Belfiore_yeff}), in the closed-box model with instant enrichment, we have
\begin{equation}
Z = \frac{y_o}{1-r}\cdot \ln(1/\mu)
\end{equation}

The effective yield defined by the equation: \begin{equation}  y_{\rm{eff}} = \frac{Z}{\ln(1/\mu)} \end{equation}differs from the true stellar yield $y_o$ by the factor $(1-r)$. Importantly, $y_{\rm{eff}}$ is independent of the star formation efficiency and history, but it may deviate significantly from $y_o$. Thus, a comparison of  $y_{\rm{eff}}$ with $y_o$ for individual galaxies can offer critical insights into the external processes that shape the chemical evolution of the interstellar medium. The true yield, $y_o$, can be estimated either from theoretical models of stellar evolution \citep[e.g.][]{Vincenzo2016}, or found by an empirical way -- for the inner parts of luminous spiral galaxies \citep{Pilyugin2007}.

To calculate the effective yield, we needed the radial profiles of the surface density of stars and gas. Here we neglected the molecular gas since its contribution is low compared to \HI\ in AGC~192040 and there are no measurements of the molecular gas content of UGC~1382. We took the \HI\ surface density profile of UGC~1382 from \cite{Hagen2016}. For AGC~192040, we derived the \HI\ surface density from the 0th moment maps from \cite{gereb2018} using the {\sc ellipse} routine. In both cases, we took into account the contribution of helium (multiplied the \HI\ surface density by 1.3).

To get the surface density of the stellar disk, we did the following. For UGC~1382, we used the {\it g}-band structural parameters of the disks given in Sect. \ref{sec:surfphot}.  The mass-to-light ratios were calculated from the $g-r$-color indices following \cite{Hagen2016}  based on the model relation from \cite{Roediger2015} \revone{after taking into account the difference between the colour in inner and outer parts of the galaxy}. However, the resulting estimates of the yield do not change significantly if we use the stellar surface density from \cite{saburovaetal2021} for this galaxy.

\begin{figure}
  \centering
  \includegraphics[width = \linewidth, trim={.1cm .25cm .4cm 0cm}, clip]{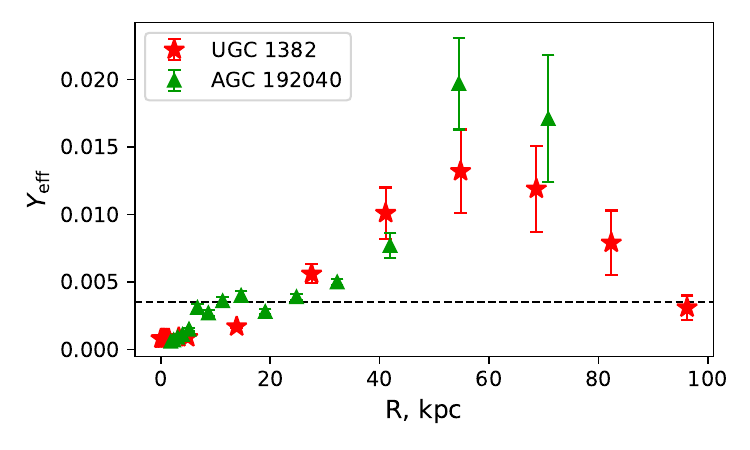}
  \caption{The radial profiles of the effective oxygen yield for UGC~1382 (stars) and AGC~192040 (triangles). The horizontal line shows $Y_o$=0.0035 obtained by \cite{Pilyugin2007} for inner regions of luminous spiral galaxies.} \label{yeff}
\end{figure}

For AGC~192040, we estimated the stellar surface density by subtracting the bulge component from the total surface brightness profile derived in Sect.~\ref{sec:surfphot}. Using our stellar population age and metallicity maps (Fig.~\ref{maps_agc}), we adopted an {\it r}-band mass-to-light ratio of 1.32. This value assumes a Salpeter IMF; adopting a Kroupa IMF or a color-based M/L estimate would lower the stellar surface densities \revone{by a factor pf $\sim 1.7$} and thus increase the inferred oxygen yield. \revone{This implies that the estimates presented here should be regarded as the lower limits of $y_{\rm{eff}}$. Note that  the systematic uncertainties related to the choice of IMF are not included in the error bars in Fig. \ref{yeff}.} 

We demonstrate the resulting radial profiles of the effective oxygen yield in Fig. \ref{yeff}. The red and green symbols represent the profiles for UGC~1382 and AGC~192040, respectively. The horizontal line indicates the value \revone{of maximal effective yield} obtained by \citet{Pilyugin2007} for the inner regions of luminous spiral galaxies, which is considered to be close to the true yield. As seen in the figure, both galaxies exhibit similar trends: within their inner regions, the effective yield is slightly lower than or comparable to that \revone{expected for } typical luminous spirals, but it increases significantly toward the outskirts, that is a relative surface density of gas becomes too high for its metallicity. \revone{The effective yield profile of UGC 1382 turns downward at the outermost point, yet it remains close to the maximum value according to \citet{Pilyugin2007}.} 
\revone{Note that for blue edge-on LSB galaxies \citet{2023ApJ...948...96C},  found the yields close to 
$y_o=0.01$ for systems with comparable stellar masses.} Thus, the evolution of chemical abundance of gas of the discussed gLSB disks is not in conflict with that of moderate-size blue LSB galaxies.  Earlier 
in \cite{Du2017}, the authors also concluded that the gas mass fraction of edge-on LSB galaxies appears to be too high for the given oxygen abundance,  but this conclusion depends on the stellar mass estimate. Note, however, that in  \citet{2004MNRAS.355..887K} the authors did not find obvious deviation from the closed box model for their sample of LSB galaxies. 

\revone{In general the high value of $y_{\rm{eff}}$ are commonly found in galaxies with large gas mass fractions \citep[see, e.g.][]{Filho2013,Ekta2010}. This trend is most evident in low-mass galaxies, but it is reasonable to expect that similar conditions may arise in regions of more massive systems - including LSB galaxies, where the gas component dominates. At the same time, such elevated effective yields cannot be attributed to an IMF enriched in massive stars, since there is evidence that low–surface-density environments instead tend to host IMFs biased toward low-mass stars \citep[e.g.][]{Lee2004, Rautio2024}.}  

The  high values of $y_{\rm{eff}}$, obtained above, and apparently relevant to other LSB disks are not necessarily indicative of enriched gas accretion.  They may also be a consequence of the non-fulfillment of the instantaneous enrichment condition, which implies the immediate, without any time delay, participation of enriched gas in the production of the next generation of stars. In the outer regions of normal galaxies and LSB galaxy disks, the contribution of the gas component to the surface density of the disk is comparable to or dominates that of stars, and star formation rates are very low. In this case, the condition of instantaneous enrichment may be unacceptable due to the lack of mechanisms for rapid gas mixing in thick, sparse gas disks with low star formation efficiency $SFE=SFR/M_{gas}$. 

 The surface density of gas $\Sigma_{gas}$ at a radial distance greater than $R > 30$ kpc prevails over the stellar density for both galaxies, being equal or lower than $3 .. 5 \Msun/$pc$^2$,  although the stellar disks with the ongoing star formation can be traced to more than 60 kpc. There are no direct estimates of the gas velocity dispersion $\sigma_g$ for the considered galaxies, but the measurements available for several galaxies with low-luminosity disks have shown that    $\sigma_g \approx   6 .. 12 $ \kms  \citep[see, for example,][]{Pickering1997,2010MNRAS.404.2061E, Eibensteineretal2023}.  Assuming $\sigma_z  \approx 8$ \kms and  $\Sigma_{gas} < 5 M_{\sun}/$pc$^2$, we obtain that the thickness of the self-gravitating gas disk: $$2h_z \approx  2\sigma_z^2/(\pi G \Sigma_{gas})$$  exceeds 2 kpc, and the volume density of gas (in the plane of the disk) is less than $2\times10^{-25} \text{g}/\text{cm}^3$. In such conditions, star formation can only occur if there are mechanisms that generate inhomogeneities in the medium, locally compressing the gas on scales smaller than the thickness of the disk, such as the low contrast density waves, the accretion of gaseous clouds, or the interaction of subhalos with the disk. Since the current SFE is very low, the predominant fraction of gas in the outer disk regions remains poorly involved in this process. This can lead to overestimates of $y_{\rm{eff}}$ obtained in a simple evolutionary model on the basis of spectral measurements of oxygen content in star-forming regions. 
 
 In other words, the apparent overabundance in the outer regions may suggest the presence of a substantial amount of ’passive’ gas -- that is a metal-poor gas, which remains weakly associated with star formation and does not contribute significantly to the emission spectrum. 

\subsection{Merger Type and Evolutionary Implications}

Our results point to mergers as the most plausible formation channel for both UGC~1382 and AGC~192040. Several lines of evidence support this view: flat metallicity gradients, systematically low gas metallicities \revone{for given stellar mass} (at least in AGC~192040\revone{, see Fig.\ref{fig:ohmstar}}), the presence of compact satellites, counter-rotation in UGC~1382, and disk warps in both systems.\footnote{We cannot fully exclude that dark matter halo properties also shaped these galaxies. However, UGC~1382 does not show signs of a "sparse" halo with high core radius and low central density \citep{saburovaetal2021}. The halo of AGC~192040 will be analyzed in a forthcoming paper.}  At the same time, neither optical images nor gas metallicity maps reveal strong merger signatures (Fig.\ref{zmap}). Since neither galaxy resides in a cluster where tidal fields might erase merger debris, such features should remain visible for up to 6–8 Gyr, depending on the mass ratio \citep{ji2014}. Thus, we can conclude that merger(s) (if they took place) occurred several Gyr ago. The settled appearance of the \HI\ disks \citep{gereb2018} is consistent with this timescale. Stellar population ages also support an early event: $\sim$6 Gyr outside the bulge region in UGC~1382 and $\sim$4 Gyr in AGC~192040, suggesting a burst of star formation triggered by a merger,  followed \revone{by a long period of quiescent evolution. Furthermore, as shown by \citet{2024ApJ...963...86L}, the relatively low-density environment in which UGC 1382  resides likely plays an important role in enabling it to acquire and retain its enormous \HI ~reservoir.}

The compact satellites associated with both galaxies may further constrain the timing of the merger. The morphology of a companion determines how quickly it is disrupted: pure bulge satellites can retain $\gtrsim$60\% of their mass after 10 Gyr, while pure disks are fully destroyed within 8 Gyr \citep{Chang2013}.  

 At the same time, we see the highly concentrated satellites, which indicates that the companion has possessed a prominent bulge. The most probable morphology of the companion that ended its evolution as the compact remnant observed in both galaxies is a composite, consisting of both bulge and gas-rich disk.

According to the simulations cited above, the composite satellite would retain up to 20 percent of its mass to $t=10$~Gyr, which does not contradict the independent age estimates above.

The other important question is related to the mass ratio of the suggested merger. The presence of regular rotating disks in both systems argues against a recent equal-mass merger. The tightly wound spiral arms that we observe in UGC~1382 and do not see in AGC~192040 are more typical for the merger mass-ratio of 6:1 than the equal mass merger \citep{ji2014}. Yet the baryonic mass ratio between the inner lenticular galaxy and the outer LSB+\HI\ disk of UGC~1382 is close to 2:1 \citep{Hagen2016}. This suggests either a major merger or multiple events that led to the tight and thin spiral structure. 

Cosmological simulations reinforce this interpretation. The EAGLE simulation shows that gLSBG analogues often undergo both major and minor mergers, with a notable excess of major (1:4 or higher) mergers relative to control sample galaxies \citep{saburova2023}. The Romulus25 simulation similarly suggests that major mergers may represent the dominant channel for moderate-size LSB galaxy formation \citep{2025arXiv250721231W}. In these cases, co-rotating companions aligned with the host galaxy’s gas disk with high orbital angular momentum enhance the host’s spin, creating diffuse, extended disks. The TNG simulations provide further support: gLSBGs can be formed in major mergers occurring 2–4 Gyr ago with preferential in-spiral orbit \citep{Zhu2023}. Notably, an analogue of Malin~1 was produced in TNG via the merger of three massive galaxies \citep{Zhuetal2018}. In TNG100, the most extreme disks are typically associated with gas-rich mergers.

UGC~1382 also hosts a stellar bar, raising the question of whether such a structure could survive or reform after a major merger. Simulations suggest that this is indeed possible. Several gLSBGs in TNG exhibit bars despite higher ex-situ stellar fractions than control samples, implying more frequent mergers. This indicates that mergers need not destroy bars permanently. Moreover, bars can be re-established after disruption \citep{Bournaud2002}. Thus, the presence of a bar in UGC~1382 does not contradict a merger-driven origin.

\subsection{Were the gLSB disks formed from the progenitors of the compact satellites?}
As discussed above, the most plausible formation scenarios for gLSBGs involve mergers. A natural question follows: could the compact satellites observed today be the primary source of material for the extended disks? To address this, we estimate the maximum stellar mass their progenitors could have contributed.

Simulations show that composite (disk+bulge) satellites can retain up to 20\% of their stellar mass after 10 Gyr \citep{Chang2013}. This implies that as much as 80\% of the progenitors’ stellar mass could have been stripped and incorporated into the host disks.

From the \revone{DESI} Legacy Survey catalog, the absolute {\it r}-band magnitudes of the satellites \revone{of UGC~1382 and AGC~192040} are $-16.3$ \revone{(C, Fig.\ref{zmap})} and $-18.7$ \revone{(B, Fig.\ref{zmap})}.
 Our spectral fitting yields the following approximate estimates of the stellar age and metallicity of the satellites: 9.1 Gyr, $-0.43$ dex \revone{(UGC~1382, C)} and 7.1 Gyr, $-0.16$ dex \revone{(AGC~192040, B)}. It gives the {\it r}-band mass-to-light ratios 3.2 and 3.26 for the Salpeter IMF in agreement with the values following from $(g-r)$ color indexes and model relation from \citet{Roediger2015}. For a Kroupa IMF, the corresponding ratios are 1.83 and 1.9. Using these ratios, we find that the progenitors could have contributed up to 1.7..2.8 $ \times 10^9$ \Msun\ to UGC~1382, and 1.6..3$ \times 10^{10}$ \Msun\ to AGC~192040.
 
It is worth comparing these values to the stellar masses of the gLSBGs disks. For AGC~192040, our photometric decomposition gives an $r$-band luminosity of $2.38\times10^{10}$ \Lsun\ for the disk+ring+spiral structure. With mass-to-light ratios of 0.77–1.32 (depending on the assumed IMF), the stellar mass is $1.83$–$3.14\times10^{10}$ \Msun. Thus, the stripped stellar material from its satellite progenitor could plausibly account for most of the stellar disk. However, the galaxy also contains a very large gas reservoir, exceeding the stellar mass, which likely required additional accretion from cosmic filaments or \revone{from the cooling hot halo gas \citep[see e.g.][]{Zhuetal2018}}.

In UGC~1382, the situation is different. The stellar mass of the spiral arms alone is \citep[1.6$\times 10^{10}$\Msun][]{Hagen2016}, which is an order of magnitude higher than the maximum contribution expected from its compact satellite. Even ignoring gas and the inner disk, a single cE progenitor cannot explain the observed disk mass. The formation of UGC~1382’s disk, therefore, likely required multiple mergers rather than the disruption of a single satellite.

Finally, there are implications for the formation of the compact satellite. According to \citet{Sharonova2025}, cEs are typically pre-processed in small groups through tidal stripping, and both UGC~1382 and AGC~192040 are members of such groups \citep{Hagen2016, gereb2018}. Moreover, massive central galaxies with masses comparable to these two systems are capable of efficiently forming cEs \citep{2010MNRAS.405L..11C, 2014MNRAS.443.1151N}. Thus, the environments of UGC~1382 and AGC~192040 provide favorable conditions for the formation of cEs through tidal stripping.

\subsection{Why do we see gaseous counter-rotation in UGC~1382 and co-rotation in AGC~192040?}

A key question concerns the contrasting kinematics of the two systems: UGC~1382 hosts a counter-rotating gaseous disk, while AGC~192040 does not. Both galaxies have compact companions that could have supplied material to their extended disks, but the orbital configuration of these companions appears to be different.

In UGC~1382, the compact satellite’s stellar velocity shares the same sign as the ionized gas, but is opposite to that of the inner stellar component. This suggests that the satellite likely followed a retrograde orbit relative to the host (inner) stellar disk, and that gas stripped from it settled into a counter-rotating configuration.

In AGC~192040, the satellite’s velocity instead aligns with both the stellar and gaseous disks, consistent with a prograde orbit. In this case, any material stripped from the companion would naturally co-rotate with the host, preventing the development of counter-rotation.

The different gas reservoirs in the two galaxies likely reinforced this outcome. AGC~192040 has a much higher gas-to-stellar mass ratio and a more massive \HI\ disk than UGC~1382. Even if some accreted material had retrograde angular momentum, it would have been overwhelmed by the dominant prograde gas component. As shown by \citet{Gasymov2025}, counter-rotating disks cannot form when the host already contains a massive gaseous disk with higher angular momentum. Additionally, the central stellar component of AGC~192040 shows no significant rotation, further suggesting a different assembly history compared to UGC~1382.

\subsection{Are UGC~1382 and AGC~192040  exceptional? }
Our results show that UGC~1382 and AGC~192040 are not complete outliers in the broader population of disk galaxies. Both follow the established relation between metallicity gradients and the exponential disk scale length seen in normal spirals. Moreover, the mode of star formation in gLSBGs resembles that of gas-rich dwarf galaxies. A similar conclusion follows from other gLSBGs mass, specific angular momentum, and gas fraction \citep{Mancera2021}.

It is worth mentioning that there also exist non-LSB galaxies with more moderate sizes, whose morphology is similar to that of gLSBGs. These are normal spiral and S0 galaxies possessing the extended gaseous disks with traces of recent or current star-formation evidenced from their UV-images --  so-called extended ultraviolet disks \citep[XUV disks][]{GildePaz2005, Thilker2005, Thilker2007, Bernaud2025}. 

Some other galaxies were initially considered typical early-type systems until \HI observations revealed the extensive gaseous disks with very low star formation activity \citep{Varnas1987,Morganti1997,2002AJ....123..729O, Serra2006, Silchenko2023}. A striking case is NGC~4203 \citep{ngc4203}, which hosts an outer star-forming \HI\ disk extending to $\sim$40 kpc. This outer component contains nine times more \HI\ than the inner galaxy. Even if all of this gas were converted into stars, its surface brightness would remain about two magnitudes fainter than the iconic gLSBG Malin~1. Such examples highlight the wide continuum of extended disk galaxies, spanning a broad range of sizes, morphologies, and surface brightness levels.

In this context, gLSBGs such as UGC~1382 and AGC~192040 may not be entirely unique. Instead, they appear to push ordinary disk galaxy processes to their extremes, offering a powerful laboratory for studying star formation, chemical evolution, and angular momentum retention in the most extended disk environments known.
\section{Summary}\label{sec:sum}

Below, we summarize the main findings and the insights of this study: 
\begin{itemize}
  \item Using our MUSE observations, we constructed detailed kinematic maps of the stellar and ionized gas components, along with spatially resolved maps of stellar age and metallicity for both stars and gas in two gLSBGs with compact satellites: UGC~1382 and AGC~192040.  We confirm gaseous counter-rotation in UGC~1382, while no counter-rotation is found in AGC~192040. Both galaxies host old, metal-rich bulges and stellar disks with mean luminosity-weighted ages of 4–6 Gyr. Their gas metallicity gradients are nearly flat. BPT diagnostics indicate the \revone{possible} presence of low-luminosity AGNs \revone{(LINERs)} in both systems.
  \item \revone{The high value of the effective oxygen yield  in the outer regions of both galaxies appears to be a consequence of the impossibility of fulfilling the condition of “instantaneous” gas enrichment, which fails in regions with  low gas density in the absence of effective mechanisms for its rapid mixing.}

  \item Despite both galaxies being gLSBGs and having compact satellites, their formation scenarios may be different. For AGC~192040, we propose the gas accretion from the filament or \revone{from the cooling hot halo gas \citep[][]{Zhuetal2018}}, followed by the intermediate mass merger with the companion possessing a prograde orbit. For UGC~1382, multiple mergers are most suitable to explain the observed properties. In both cases, the systems have had enough time to settle down -- the last mergers responsible for their observed gas properties should happened  $6-8$ Gyr ago.
\end{itemize}

\begin{acknowledgements}
\revone{The authors thank the anonymous referee for helpful and encouraging comments. }The authors thank Barbara Catinella for kindly providing the \HI\ data cube and moment maps of AGC~192040.
Based on observations collected at the European Southern Observatory under ESO program 110.24DN (PI: Saburova)
  AZ, IG, DG,  ER research on the analysis and fitting of the spectral data, gas metallicity estimates was supported by The Russian Science Foundation (RSCF) grants No.~23-12-00146. IC's research is supported by the Telescope Data Center at Smithsonian Astrophysical Observatory. OE acknowledges funding from the Deutsche Forschungsgemeinschaft (DFG, German Research Foundation) -- project-ID 541068876.
  The Legacy Surveys consist of three individual and complementary projects: the Dark Energy Camera Legacy Survey (DECaLS; Proposal ID 2014B-0404; PIs: David Schlegel and Arjun Dey), the Beijing-Arizona Sky Survey (BASS; NOAO Prop. ID 2015A-0801; PIs: Zhou Xu and Xiaohui Fan), and the Mayall z-band Legacy Survey (MzLS; Prop. ID 2016A-0453; PI: Arjun Dey). DECaLS, BASS and MzLS together include data obtained, respectively, at the Blanco telescope, Cerro Tololo Inter-American Observatory, NSF’s NOIRLab; the Bok telescope, Steward Observatory, University of Arizona; and the Mayall telescope, Kitt Peak National Observatory, NOIRLab. Pipeline processing and data analyses were supported by NOIRLab and the Lawrence Berkeley National Laboratory (LBNL). The Legacy Surveys project is honored to be permitted to conduct astronomical research on Iolkam Du’ag (Kitt Peak), a mountain with particular significance to the Tohono O’odham Nation.
\end{acknowledgements}

\bibliographystyle{aasjournalv7} 
\bibliography{LSB} 

\end{document}